\documentclass[twocolumn,superscriptaddress,aps]{revtex4}

\usepackage{amsmath,amsthm,amsfonts}
\usepackage{epsfig}
\usepackage{graphicx}
\usepackage{amssymb}
\usepackage{color}      
\usepackage{natbib}
\usepackage[version=3]{mhchem}
\newcommand{\Eq}[1]{Eq. (\ref{#1})}

\newcommand{\ket}[1]{| #1 \rangle}

\newcommand{\beq}{\begin{eqnarray}}
\newcommand{\eeq}{\end{eqnarray}}

\begin{document}
\title{Radical-pair model of magnetoreception with spin-orbit coupling}
\author{Neill Lambert}
\affiliation{CEMS, RIKEN, Saitama, 351-0198, Japan}
\author{Simone De Liberato}
\affiliation{ School of Physics and Astronomy, University of Southampton, Southampton, SO17 1BJ, UK}
\author{Clive Emary}
\affiliation{Department of Physics and Mathematics, University of
Hull, Hull HU6 7RX, SUnited Kingdom}
\author{Franco Nori}
\affiliation{CEMS, RIKEN, Saitama, 351-0198, Japan}
\affiliation{Department of Physics, University of Michigan, Ann Arbor, Michigan
48109-1040 USA}
\bibliographystyle{apsrev}
\begin{abstract}
The mechanism used by migratory birds to orientate themselves
using the geomagnetic field is still a mystery in many species.
The radical pair mechanism, in which very weak magnetic fields can
influence certain types of spin-dependent chemical reactions,
leading to biologically observable signals, has recently imposed
itself as one of the most promising candidates for certain
species. This is thanks both to its extreme sensitivity and its
capacity to reproduce results from behavioral studies. Still, in
order to gain a directional sensitivity, an anisotropic mechanism
is needed. Recent proposals have explored the possibility that
such an anisotropy is due to the electron-nucleus hyperfine
interaction. In this work we explore a different possibility, in
which the
 anisotropy is due to spin-orbit coupling between the electron spin and its angular momentum.
We will show how a spin-orbit-coupling-based magnetic compass can
have performances comparable with the usually-studied
nuclear-hyperfine based mechanism. Our results could thus help
researchers actively looking for candidate biological molecules
which may host magnetoreceptive functions,  both to describe
magnetoreception in birds as well as to develop artificial
chemical compass systems.
\end{abstract}

\maketitle


Growing evidence suggests  quantum  coherence plays an important
role in hitherto unexplained biological phenomena
\cite{Lambert12}. The two most widely known examples are the
recent observation of coherent energy transport in photosynthetic
bacteria, that could help explaining the almost unitary efficiency
of energy harvesting \cite{EngelNature07,Lambert12} and the role
of coherence in the navigation ability of migrating birds
 \cite{Lambert12,Wiltschko66,Schulten,Hore2012}.

It has been proposed \cite{Schulten,Ritz00,Ritz10} that electron
spin could act as a detector for the Earth's geo-magnetic field
(of magnitude $0.5$ G $\equiv 50$ $\mu$T). The mechanism, called
the radical pair mechanism (RPM), works as follows: upon
absorption of a photon, a molecule (or part of a molecule, in
cryptochrome for example) is photochemically excited, to form a
radical pair with another nearby molecule (an electron is
transferred to the excited acceptor from a donor molecule). This
radical pair inherits its singlet or triplet nature from the
molecular precursors. The Earth's magnetic field then causes
oscillations of the electron spins. As radical pairs have a finite
lifetime, and certain biological processes could be sensitive to
the difference in metastable chemical products that result
depending on the singlet or triplet nature of the radical pair at
the time of decay, it has been proposed that the whole system may
then be able transduce information of the Earth's magnetic field
direction into chemical signals that the bird can perceive.

The RPM model, while originally proposed as a possible explanation
for magnetoreception several decades ago \cite{Schulten}, recently
gained renewed attention because it can explain some of the
results of recent behavioral experiments.  In these experiments
the application of radio frequency fields \cite{Ritz04,Ritz09} or
the alteration of ambient light conditions
\cite{Wiltschko00,WiltLight} disrupted avian navigation. The
radical pair model has also been strengthened by the discovery of
a candidate radical pair system with the photo-receptor
cryptochrome
\cite{Maeda2012,Ilia07,Moller04,Mouritsen04,Liedvogel07,liedvogel10}.
In addition, neuroanatomical evidence \cite{Zapka09} strongly
suggests that a vision-mediated pathway underlies magnetoreception
in European robins, further strengthening the notion that
photo-activated \cite{Wiltschko00,WiltLight} radical pairs
resident in the
 eye could be the mechanism underlying a certain type of magnetoreception (recent evidence strongly suggests there may several different mechanisms at play \cite{Hore2012}).

 However, the radical pair model still faces some challenges. As yet,
sufficient sensitivity to external fields has yet to be confirmed
in \textit{in vitro} experiments on candidate radical pairs.  This
is particularly demanding in terms of describing the disruption of
the navigational sense of birds who are exposed to oscillating
magnetic fields \cite{WiltLight, Ritz04,Ritz09, Wiltschko07} as
weak as $15$ nT.  Gauger {\it et al.}~\cite{vedral11} showed that
while the compass itself is immune to certain types of phase noise
(and thus can, in some sense, operate in a classical limit),
coherence must be retained for $10$s of $\mu$s, and ideally more
than 100 $\mu$s, to explain this disruptive effect. A recent
analysis found that the interaction of the electron spins with a
spin-bath can stabilize coherence and also provide another
alternative route to achieve sensitivity to the angle of the
external field \cite{Walters}. Finally, the RPM model demands some
sort of {\em anisotropic} interaction to be
directionally-sensitive, and requires some order in the alignment
of radical pairs in at least one axis \cite{Hill10, Ilia10,Lau2},
though recent work suggests, because of the selective way
radical's may be photo-excited, ordering may be unnecessary
\cite{Lau,Hore2012}.

Most of the discussion in this topic is based on the same
assumption that the anisotropy is given by the nuclear spin
hyperfine interaction \cite{Schulten, Ritz04}. Still, this is only
one of the possible mechanisms that could lead to the needed
anisotropic interaction. In this paper we investigate the role of
another candidate mechanism, namely the spin-orbit coupling (SOC)
in radical pair systems. Depending on the type of radical pair,
SOC can either cause anisotropic $g$-factors (in distant radical
pairs) or spin-selective transition rates (in close, or contact,
radical pairs); in both cases producing a level of anisotropy
suitable for a chemical compass to operate.

Anisotropic g-factors have been discussed elsewhere as a possible
source of magnetic field sensitivity \cite{Hogben1,vedral11,
Erik,Hogben2}.  In particular, Ref.~\cite{Hogben1} critically
discussed SOC in cryptochromes when either dioxygen or superoxide
form one half of the radical pair.  The advantage of this
assumption is that oxygen has little nuclear hyperfine
interaction,  thus enabling the combination of oxygen with the
flavin adenine dinucleotide (FAD) cofactor in cryptochrome to
realize the highly anisotropic nuclear spin model needed to
explain the disruptive effect of external oscillating fields (the
so-called Zeeman resonance). However, they argued that SOC in
radical pairs involving oxygen in solution lead to rapid loss of
spin-coherence due to periodic unquenching of the SOC. If the
radicals are immobilised in a lattice, this effect could be
suppressed. However, the strong nuclear hyperfine interaction in
FAD would still dominate over anisotropy from the SOC, and hence
the normal nuclear hyperfine model would probably apply.  In this
work we consider both the ideal situation where the SOC dominates,
and in the final section when both SOC and hyperfine interactions
play a role.

The rest of this paper is organized as follows: after having
introduced, in Section \ref{RPM}, the general features of a RPM,
we will summarize, in Section \ref{HYP}, the results of
Ref.~\cite{vedral11} for a RPM based on nuclear spin interaction,
that has been extensively studied in the literature and which has
been termed the ``reference probe model''. In Sections \ref{SOC1}
and \ref{SOC2} we will then investigate the efficiency of
SOC-based RPM. In Section \ref{SOC1} we will see in what ways a
SOC-induced anisotropic $g$-factor can cause an efficient response
to external fields.  Of particular interest is the fact that very
weak $\Delta g$ effects (as are expected in organic radical pairs)
can have a large magnetic-field sensitivity if the lifetime
approaches $100$ $\mu$s, giving a possible practical explanation
for why the radical-pairs may need such a long lifetime
\cite{Erik}. We emphasize that this is not a limiting factor in
the SOC model, when taken in context with the standard model of
radical-pair reactions, as, when the oscillating field behavioral
experiments \cite{Ritz04,Ritz09} are taken into account, these
standard models also demand either such a long life-time
\cite{vedral11} or that the magneto-reception mechanism is
extremely sensitive to minute changes in product yield.  The
alternative is that some different models or mechanisms may arise
which remove or reduce this demand \cite{Hore2012}. In Section
\ref{SOC2} we discuss the effects of SOC-induced spin-dependent
recombination rates that are found in contact radical pairs. We
will conclude in Section \ref{CON} with a comparison of the
various mechanisms and suggestions for future investigations.

\section{Results}
\subsection{General framework of the radical pair mechanism} \label{RPM} In the
radical pair mechanism
 \cite{Schulten,Ritz00}  a radical pair is formed due to an optical
excitation process and the associated transfer of an electron from
a donor to an acceptor molecule. The excess electrons retain their
initial correlation (which may be assumed to be a singlet, but can
also be a triplet state without a significant change in the
traditional RPM effect).

The correlated spins then evolve under the
 joint influence of the geomagnetic field $\boldsymbol{B}$ and of other magnetic interactions (e.g., nuclear spins,  molecular fields, or orbital momentum).
If the evolution of the two spins is different, e.g., because one
of the additional interactions is in-homogeneous (e.g., if only
one of the radicals has appreciable nuclear hyperfine
interactions), it will cause an oscillation between singlet and
triplet states. Eventually the electrons will undergo a
spin-selective recombination. The yield of the reaction is then
sensitive to the dynamics of the two spins, and it can be used to
measure the direction of $\boldsymbol{B}$. 

The effect of the external magnetic field on the electrons is described by the Hamiltonian
\begin{equation}
\label{HB} H_{B}=
\frac{1}{2}\mu_{\rm{B}}(\boldsymbol{B}\cdot\bar{g}_1\cdot\hat{\sigma}_{1}+\boldsymbol{B}\cdot\bar{g}_2\cdot\hat{\sigma}_{2}),
\end{equation}
where  $\bar{g}_1$ and $\bar{g}_2$ are tensors describing the
generally anisotropic $g$-factors. For clarity we have used Pauli
spin matrices here, hence the factor of $1/2$. To simplify
matters, an axial symmetry is usually assumed; then the magnetic
field vector can be written as (including a possible external
time-dependent contribution)
$\boldsymbol{B}=B_{0}(\sin\theta,0,\cos\theta)+
B_{\rm{rf}}\cos\omega_d t(\sin\theta',0,\cos\theta')$, for
$\theta,\theta' \in[0,\pi/2]$.

In addition, the radical-pair is affected by several decoherence
channels. The primary channel is spin-selective
recombination into the singlet and triplet product states
$\left|\textbf{s}\right\rangle$ and
$\left|\textbf{t}\right\rangle$. This occurs when an excess
electron returns from the acceptor radical to the donor radical,
and can be generally  described in the Lindblad form by the
superoperator
\begin{equation}
\Sigma[\rho]=-\sum_{\alpha}\kappa_{\alpha}\left[\frac{1}{2}s_{\alpha}s_{\alpha}^{\dag}\rho-s_{\alpha}^{\dag}\rho
s_{\alpha}+\frac{1}{2}\rho
s_{\alpha}s_{\alpha}^{\dag}\right],\label{gammar}
\end{equation}
where the $s_{\alpha}$ are jump operators describing the model under examination.
Additional dephasing and amplitude noises can be included in the model \cite{vedral11}, which may arise from
sources like dipole interactions, magnetic fluctuators in the biological environment, and so on.  We will discuss this in the final section.

One possible biologically-detectable signal (though others have
recently been proposed \cite{Erik}) is the triplet or singlet
yield. For example, in the normal Liouville treatment the singlet
yield is \beq \Phi(\theta) = \kappa \int_0^{\infty} S(t)\; dt,\eeq where
$S(t) = \mathrm{Tr}[Q^S\rho(t)]$, $\kappa$ is the radical pair decay rate, and $Q^S$ is the projector onto
the singlet states.   Here, we solve the master equation,
$\dot{\rho}(t)=\mathcal{L}[\rho(t)]$, where $\mathcal{L}$ is
$\mathcal{L}=-i[H,\cdot]+\Sigma$, and simply observe the
occupation of the ``shelving'' state $\ket{\mathbf{s}}$, which in
the long-time limit is equivalent to the singlet yield.
  For a given compass model to be deemed as sufficiently sensitive
  one needs a large change in $\Phi(\theta)$ as a function of $\theta$. Thus often the quantity of interest
  is the sensitivity   \beq \Phi_{\rm{D}} =
\Phi(\pi/2) -\Phi(0), \quad |\Phi_{\rm{D}}| \leq 1.
\label{sens}\eeq In the following Sections we will calculate the
sensitivity $\Phi_{\rm{D}}$ both for static and oscillating
magnetic fields, for three different anisotropy-inducing
mechanisms (where, in absence of any anisotropy,
$\Phi_{\rm{D}}=0$).

\subsection{Radical pair mechanism with nuclear-hyperfine interaction} \label{HYP}
\subsubsection{Hyperfine interaction}

The nuclear hyperfine interaction (assuming only one of the
electrons is significantly affected by the nuclear spin
\cite{Ritz10, Christopher09}) is described by,
\begin{equation}
\label{HHyp}
H_{\text{Hyp}}= \hat{\sigma}_{1}\cdot \frac{\mathbf{A}}{2} \cdot \hat{\sigma_I},
\end{equation}
where $\hat{\sigma_I}$ is the nuclear spin operator, assumed for
simplicity to be a single spin such that $\sigma_I$ is a Pauli
spin operator, and $\mathbf{A}=\mathrm{diag}(A_{x},A_{y},A_{z})$
is the anisotropic hyperfine coupling tensor. The factor of $1/2$
allows us to directly compare the magnitude of $A/g\mu_{\rm{B}}$
to $B_0$ in units of $\mu$T. We also omit an additional factor of
$1/2$, normally associated with the nuclear spin operator, to
enable this comparison. This has been termed the reference-probe
model \cite{Ritz10}, as it represents an extremely simple example
with just a single nuclear spin degree of freedom, which produces
a very sensitive compass against which more complex real models of
radical pairs can be tested.

The spins thus evolve under the conjoint effect of the
Hamiltonians in \Eq{HHyp} and \Eq{HB}, that now takes the form (setting the g-factor as isotropic, $g=2$),
\begin{equation}
H_{B}=\mu_{\rm{B}}\,\boldsymbol{B}\cdot
(\hat{\sigma}_{1}+\hat{\sigma}_{2}).
\end{equation}

The Lindblad superoperator describing spin-selective recombination, and taking into account the nuclear spin states, now reads
\begin{equation}
\Sigma_{\text{Hyp}}[\rho]=-\kappa\sum_{\alpha,\beta}\left[\frac{1}{2}s_{\alpha\beta}s_{\alpha\beta}^{\dag}\rho-s_{\alpha\beta}^{\dag}\rho
s_{\alpha\beta}+\frac{1}{2}\rho
s_{\alpha\beta}s_{\alpha\beta}^{\dag}\right],\label{gammar2}
\end{equation}
where $\alpha=1,2,3,4$ indexes the electronic spin states,
$\beta=\uparrow,\downarrow$ the nuclear spin states and \beq
s_{1\beta}&=&\left|S,\beta\right\rangle\,\left\langle
\textbf{s}\right|,\quad\quad
s_{2\beta}=\left|T_{+},\beta\right\rangle\,\left\langle
\textbf{t}\right|,\\
s_{3\beta}&=&\left|T_{0},\beta\right\rangle\,\left\langle
\textbf{t}\right|,\quad\quad
s_{4\beta}=\left|T_{-},\beta\right\rangle\,\left\langle
\textbf{t}\right|. \eeq We assume that the radical pairs in the
singlet and triplet states have the same recombination rate, or
decay rate, $\kappa$.

\subsubsection{Results for static fields}

The hyperfine coupling tensor was found \cite{vedral11} to give a
large sensitivity when $A_{x}=A_{y}=A_{z}/2$, and
$A_{z}/g\mu_{\rm{B}} \gg B_0$, hence the nuclear spin acts like a
strong reference field.  For $B_0=47$ $\mu$T,
$A_z/g\mu_{\rm{B}}=174$ $\mu$T, one obtains a sensitivity of
$\Phi_{\rm{D}} \approx 0.1$ (see figure \ref{figure1}(a)). For
weaker field strengths we see several sharp resonances, as the
hyperfine coupling becomes comparable to the geomagnetic field.
The width of these resonances are strongly dependent on the rate
$\kappa$, and may not be useful for magnetoreception as they would
be susceptible to small changes in the nuclear hyperfine coupling
strength and $\kappa$ \cite{vedral11}).

Recently, Cai \textit{et al.}~\cite{Cai12} showed that an
optimization of the RPM, over all parameters
 in the reference-probe model, gives a maximum sensitivity of $\Phi_{\rm{D}} \approx 0.4$.  This maximum occurs for a highly
an-isotropic  hyperfine tensor of $A_{x}=A_{y}=0$ and
$A_{z}/g\mu_{\rm{B}}\approx B_0/6$ (in our definition of the
nuclear hyperfine Hamiltonian, which differs from Cai \textit{et
al.}~by a factor of 1/2). However, as with the resonances in
figure \ref{figure1}(a), this optimal point changes as a function
of $\kappa$. A similar maximum is shown in figure
\ref{figure1}(b), for a highly an-isotropic tensor.

\subsubsection{Results for oscillating fields}

The primary evidence for the RPM stems from its ability to explain
behavioral experiments on certain avian species.  This includes
photo-sensitivity: the test subjects could not navigate when
deprived of certain frequencies of light
\cite{Wiltschko00,WiltLight}.  It also includes disruption of the
navigation sense under the influence of very weak ($15$--$150$ nT)
radio-frequency oscillating magnetic fields \cite{Ritz04,Ritz09}.
In the model we described earlier this is represented by the term
$B_{\mathrm{rf}}$. One can conservatively choose $B_{\mathrm{rf}}
=150$ nT, to match the experimental conditions (though disruption
was seen also for much smaller field magnitudes),  $\omega_d =
\mu_{\rm{B}} g B_0/\hbar$ , which for $B_0=47$ $\mu$T gives
$\omega_d/2\pi = 1.32$ MHz, and in all of the following results we
set $\theta'=\theta+\pi/2$, as this maximizes the effect of the
time-dependent contribution. This induces Rabi-oscillations in the
``free'' electron state at a frequency proportional to the field
strength $B_{\rm{rf}}$.  This is an extremely slow frequency
compared to the other system parameters, and for it to have any
effect on the singlet yield it is immediately clear that $\kappa$
must be exceptionally slow. For any non-negligible effect on the
singlet yield \cite{vedral11} one needs a rate $\kappa < 10^4$
s$^{-1}$, which corresponds to a lifetime exceeding $100$ $\mu$s.
This is shown in figure \ref{figure2}(a), where the yield is
unaffected by the oscillating field unless $\kappa$ is
exceptionally small. Now that we have set the stage with the
traditional nuclear hyperfine model of the radical-pair compass,
we will next outline how spin-orbit coupling can fulfill a similar
role.


\subsection{Radical pair mechanism with spin-orbit coupling in distantly bound radical pairs}
\label{SOC1}

\subsubsection{Spin-orbit coupling and anisotropic $g$-factor} The Hamiltonian
responsible for SOC is of the form \cite{Khud}
\begin{eqnarray}
\label{HSOC}
H_{\text{SOC}}&=&\sum_j \zeta_j \mathbf{L}_j\mathbf{S}_j,
\end{eqnarray}
where $\mathbf{L}_j$ and  $\mathbf{S}_j$ are respectively the  orbital momentum and the spin operator of electron $j$ and the sum is over the valence electrons.
The value  $\zeta_j$ is termed the SOC constant and it includes the screening effect due to the presence of inner electrons.
The effect of the Hamiltonian in \Eq{HSOC} depends upon the symmetry of the system. In atoms the valence orbitals along different axes (e.g., $P_x$, $P_y$ and $P_z$) are degenerate and transitions between them thus couple with the spin degree of freedom. In molecules, this degeneracy is usually lifted, and the SOC effect can thus be quenched
 (its strength depends in this case on the ratio between the SOC constant and the splitting between the orbitals).
Of particular interest for us is the case in which the degeneracy
is only partially lifted, as in axial molecules, in which the two
orbitals orthogonal to the molecular axis (that we will assume to
be $z$) are still degenerate. In these molecules the SOC effect
will thus be quenched in the $x$--$y$ plane but present along the
$z$ direction, leading to an anisotropic $g$-factor. Physically we
can visualize this phenomenon as a current that flows around the
$z$ axis, due to the fact that the unpaired electron can freely
jump between the degenerate $x$ and $y$ orbitals. This current
will then create a magnetic field along the $z$ axis, that will
couple with the $z$ component of the electron spin.

The interaction of the radical pair with an external magnetic
field will thus be described by the Hamiltonian with an
anisotropic $g$-factor. According to the reference-probe model
logic, we will study the sensitivity of the system using, in
\Eq{HB}, a $g$-factor of the form
\begin{eqnarray}
\label{g-ani}
\bar{g}_1 &=& 2\nonumber \\
\bar{g}_2 &=& (2,2,2+\Delta g),
\end{eqnarray}
that is, only the second radical has a significant $g$-factor anisotropy.

In most organic molecules such effects are quenched ($\Delta g
\approx 10^{-2}-10^{-3}$), but in simpler radical pairs
\cite{Hogben1}, or radicals containing heavy atoms, they can be
extremely large ($\Delta g \approx 2$). The two spins will thus
evolve under the effect of \Eq{HB} with the anisotropic $g$-factor
defined in \Eq{g-ani} and they will be subject to a spin-selective
recombination, analogous to the case of the hyperfine interaction,
described in the Lindblad form as
\begin{equation}
\label{SigmaSOC}
\Sigma_{\text{SOC}}[\rho]=-\kappa\sum_{\alpha}\left[\frac{1}{2}s_{\alpha}s_{\alpha}^{\dag}\rho-s_{\alpha}^{\dag}\rho
s_{\alpha}+\frac{1}{2}\rho s_{\alpha}s_{\alpha}^{\dag}\right],
\end{equation}
where $\alpha=1,2,3,4$, $s_{1}=\left|S\right\rangle\,\left\langle
\textbf{s}\right|$, $s_{2}=\left|T_{+}\right\rangle\,\left\langle
\textbf{t}\right|$, $s_{3}=\left|T_{0}\right\rangle\,\left\langle
\textbf{t}\right|$, and
$s_{4}=\left|T_{-}\right\rangle\,\left\langle \textbf{t}\right|$.

\subsubsection{Results for static fields} 
\begin{widetext}

\begin{figure}[tbp]
    \includegraphics[width=\columnwidth]{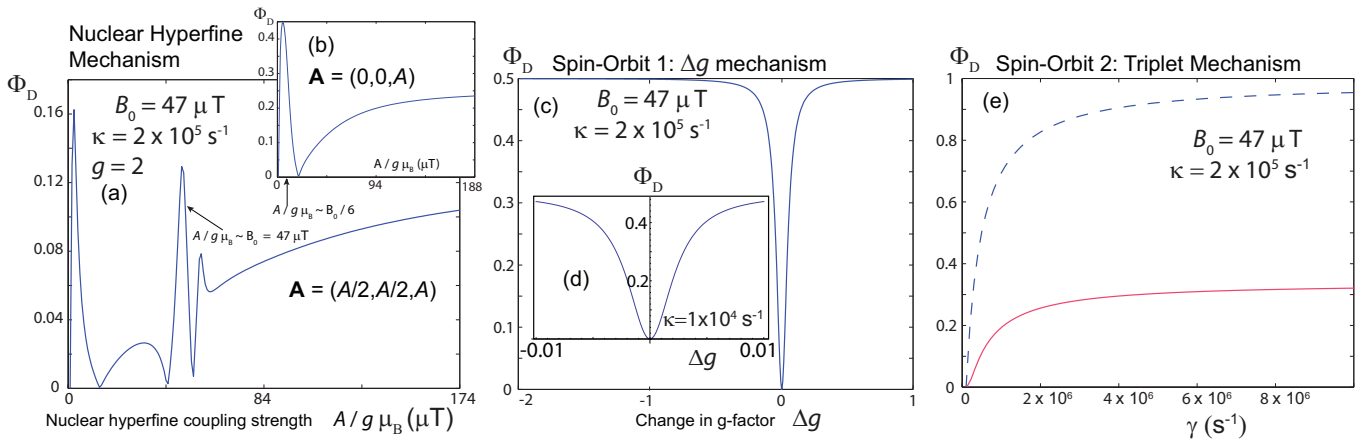}
    \caption{(Color online)
    The sensitivity of all three radical-pair mechanisms, as a
function of the appropriate parameter in each case.  The values of
the other parameters are indicated on the figures.  For (a) and
(b), the nuclear hyperfine mechanism, the sensitivity is shown as
a function of the strength of the coupling to the nuclear spin.
(a) shows the just slightly anisotropic tensor, while (b) is the
fully anisotropic tensor.  The largest sensitivity, as discussed
by Cai {\em et al.}~\cite{Cai12}, appears for
$A/g\mu_{\rm{B}}\approx B_0/6$, but is slightly shifted due to the
different value of $\kappa$ employed here. Note that the
resonances in (a) and (b) for small and intermediate values of $A$
become sharper, and some are shifted, if $\kappa$ is reduced.
Figures (c) and (d) show the SOC $\Delta g$ mechanism sensitivity
as a function of $\Delta g$. Figure (c) is for the same $\kappa$
as figure (a) whereas (d) is for a much longer-lived radical, and
thus a much smaller $\kappa$.  Such long-lived radicals are very
sensitive to even small changes in g-factor. (e) shows the
sensitivity for the triplet mechanism, as a function of the
singlet-triplet mixing rates $\gamma=\gamma_+=\gamma_-$, for an
initial pure triplet state $T_0$ (blue dashed curve) and a full
mixture of all three triplet states (red curve).
 \label{figure1}
 }
\end{figure}
\end{widetext}

In order to measure the
efficiency of the anisotropic $g$-factor as compass we solve the
full numerical model described above. We have also compared our
numerical results to the analytical results from the Liouville
equation approach in Ref.~\cite{Cai12} (which is easily adapted to
include the anisotropy in the $g$-factor). In Figure
\ref{figure1}(c) we show the sensitivity $\Phi_{\rm{D}}$, as a
function of $\Delta g$ for $\kappa = 2\times 10^5$ s$^{-1}$, and
in (d) $\kappa = 1\times 10^4$ s$^{-1}$. We find that in Figure
1(c) a large sensitivity $\Phi_{\rm{D}}=0.5$ arises for $|\Delta
g|>0.5$.

Interestingly, if $|\Delta g|$ is small, one needs a smaller
$\kappa$ to saturate the sensitivity.  For $\kappa = 1\times 10^4$
s$^{-1}$, as shown in Figure 1(d), the magnetic sensitivity is
saturated for values as small as  $|\Delta g|\approx  0.01$. One
could argue that this could explain the need for the exceptionally
long lifetime of radical pairs. In the nuclear hyperfine case, a
long lifetime is arguably unnecessary to saturate the compass
sensitivity, but is needed to explain the disruption effects of
weak oscillating fields.  Of course, this disruptive effect is not
a functional necessity of the compass, so the difficulty is in
explaining why the radical pairs would have a long lifetime which
is functionally not useful. An alternative scheme was recently
proposed \cite{Erik} to give this long lifetime (and,
incidentally, coherence time) a functional role. The dependence we
show here, of the sensitivity of a SOC mechanism with a  small
$\Delta g$ on the long life-time of the radical pair, could, in
principle, also explain this phenomenon.

\subsubsection{Results for oscillating fields}

As with the nuclear hyperfine mechanism, the SOC must also be able
to explain the rather demanding results of the behavioral
experiments. The behavior of a general $\Delta g$ mechanism under
oscillating fields was presented in the supplementary information
of Ref.~\cite{vedral11}, albeit without reference to the source of
the anisotropy.  Here we make a similar examination and find that
for large $\Delta g$, as with the nuclear hyperfine field, that
the SOC-compass is only sensitive to the external field for very
small $\kappa$ [see figure \ref{figure2}(b)].  For the interesting
case of a small shift $\Delta g=-0.01$, we show in figure
\ref{figure2}(c) that the effect of the oscillating field is
different from the large $\Delta g$ and nuclear hyperfine case. As
$\kappa$ becomes small the sensitivity is increased in the
presence of the oscillating field.  Arguably this could still
cause a disruptive behavioral effect, depending in exactly what
way the singlet yield is transduced into a directional signal by
the avian host.

A similar argument was recently made in Ref.~\cite{ban}, where
they noted that the application of additional weak static fields,
causing disruption of the magnetic sense, does not reduce the
sensitivity of the normal reference-probe model, but shifts the
overall `curve', implying there is a `window' of operation.  With
static fields, it was observed that the disruptive effect was
temporary, and eventually the birds adjusted and were once more
able to navigate \cite{wiltstatic,Ritz2011}.  If very weak
anisotropic g-factors, mediated by SOC, is the cause of magnetic
sensitivity in these species, then a similar argument may be
applicable.  The effect of oscillating fields would not suppress
the sensitivity, as in the nuclear hyperfine or strong $\Delta g$
mechanisms, but would shift the sensitivity curve out of an
operational window. Of course, if such an effect is directly
analogous to that of static fields, then also the disruptive
effect should be temporary.


\subsection{Radical pair mechanism with spin-orbit coupling in contact radical-pairs} \label{SOC2}

\subsubsection{Triplet mechanism} The second possible SOC effect is
the so-called triplet mechanism \cite{oldpaper} that can occur in
closely bound, or contact, radical pairs. As in this case the pair
can be considered as a single system, the SOC Hamiltonian in
\Eq{HSOC} can give rise to transitions involving at the same time
a variation of the orbital and spin  degrees of freedom. This
results in an anisotropic, incoherent transition from the triplet
states to the singlet ground state. For our purposes we model this
as follows:  unlike in the previous cases, we assume the radical
pair is created either in one of the triplet states or an equal
mixture of all the triplet states. The triplet states are then
coupled to each other because of the external magnetic field and
in turn undergo their normal recombination into the triplet
product state at a rate $\kappa$, as before. The effect of the
spin-orbit interaction then in this case is to induce an
incoherent inter-system crossing transition from the triplet
states to the singlet recombination product or ground state, the
rate or frequency of which is dependent on the original triplet
state. Since the relative triplet state occupations depend on the
orientation of the external field, this creates, as with the
previous mechanisms, a singlet state product which is sensitive to
the angle of the external field.

We include this in the standard radical pair model, akin to the
approach taken in Ref. \cite{oldpaper}, by allowing for
spin-selective transitions from the triplet states to the singlet
product. The Lindblad superoperator thus reads,
\begin{equation}
\Sigma_{\text{Tot}}=\Sigma_{\text{SOC}}+\Sigma_{\text{TM}},
\end{equation}
where $\Sigma_{\text{SOC}}$ is defined in \Eq{SigmaSOC} and describes the usual
 channels and $\Sigma_{\text{TM}}$ is given by
\begin{equation}
\Sigma_{\text{TM}}[\rho]=-\sum_{\alpha}\gamma_{\alpha}\left[
\frac{1}{2}s_{\alpha}s_{\alpha}^{\dag}\rho-s_{\alpha}^{\dag}\rho
s_{\alpha}+\frac{1}{2}\rho
s_{\alpha}s_{\alpha}^{\dag}\right],\label{gammar3}
\end{equation}
with  $\alpha=-,0,+$,
 $s_{+}=\left|T_+\right\rangle\,\left\langle
\textbf{s}\right|$, $s_{-}=\left|T_{-}\right\rangle\,\left\langle
\textbf{s}\right|$ and $s_{0}=\left|T_{0}\right\rangle\,\left\langle
\textbf{s}\right|$. Note in this case the g-factors are now isotropic.

\subsubsection{Results for static fields}

Generally, any anisotropic choice of $\gamma_{\alpha}$ rates will
introduce an angular dependence.  We follow the example of Ref.
\cite{oldpaper} and choose $\gamma_0=0$, $\gamma_+=\gamma_-=\gamma
$. The exact dependence or form of these parameters
 will depend on the microscopic features of the given radical, thus this choice is again
 in the spirit of an ideal `reference-probe' model. In figure \ref{figure1}(e) we plot
 the sensitivity as a function of $\gamma$.  The red lower curve shows the sensitivity for an initial complete mixture of the three triplet
states, while the blue curve shows the sensitivity for an initial
pure triplet state $T_0$, which produces an overall much larger
variation in magnitude.

\subsubsection{Results for oscillating fields}

\begin{widetext}

\begin{figure}[tbp]
    \includegraphics[width=\columnwidth]{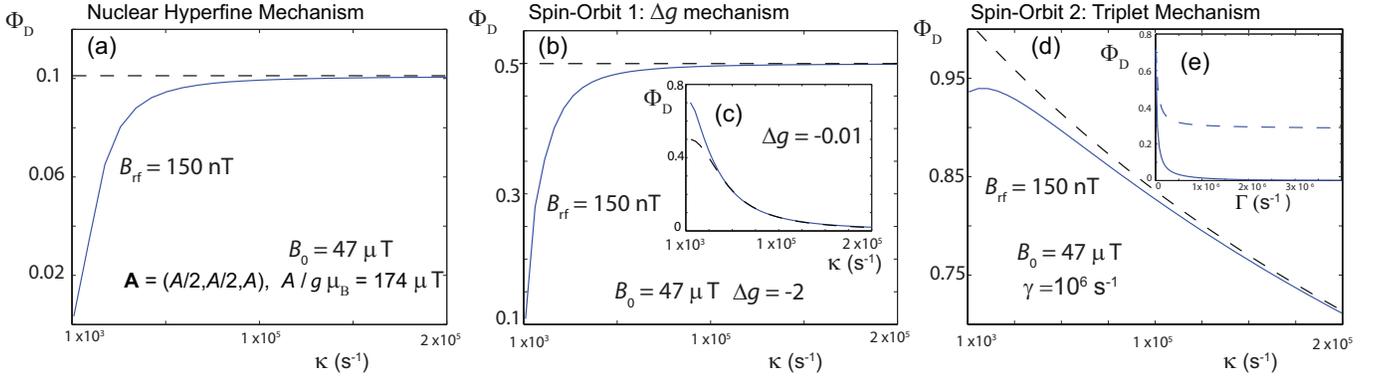}
    \caption{(Color online)
The sensitivity $\Phi_{\rm{D}}$ as a function of recombination
rate $\kappa$,
 with an external field strength $B_0 = 47$ $\mu$T, and both $B_{\mathrm{rf}}=0$ (black dashed lines) and $B_{\mathrm{rf}}=150$
nT  (blue curves). In all cases $\kappa$ was cut off at a lower
value of $1\times 10^{3}$ s$^{-1}$. (a) is for a strong nuclear
hyperfine coupling $A_z/g\mu_{\rm{B}}=174$ $\mu$T, with a
semi-anisotropic tensor, which exhibits an almost complete loss as
$\kappa\rightarrow 10^{3}$ s$^{-1}$ (these results are identical
to those in \cite{vedral11}).  (b) shows the sensitivity for the
SOC $\Delta g$ mechanism with $\Delta g=-2$. The disruption due to
the oscillating field is similar to the hyperfine one, with
significant changes in the sensitivity only appearing for
remarkably small $\kappa$.  (c) shows the effect of the
oscillating field for $\Delta g=-0.01$. In contrast to the
previous examples, the sensitivity is increased when the
oscillating field is switched on. More precisely, and not shown in
the figures, for small $\kappa$ switching on the oscillating field
reduces the yield of $\Phi(0)$ while that of $\Phi(\pi/2)$ is
generally unaffected.  Arguably this could also lead to a
disruption of the magnetic sense, though in a different way than
the `normal' flattening of the $\Phi(\theta)$ profile
\cite{vedral11}. (d) shows the sensitivity for the triplet
mechanism, for a pure initial triplet state. In this case the
overall behavior is slightly different from the hyperfine and
$\Delta g$ mechanisms, with the overall sensitivity increasing as
$\kappa$ is decreased, and the rf-field having less of a profound
effect on the magnitude.  The inset (e) shows the effect of an
additional dephasing and decoherence on the triplet mechanism. The
solid line is for six equal additional decoherence operators (see
main text), while the dashed line is just dephasing in the
$\sigma_z$ basis, and in both cases $\kappa=2\times10^5$ s$^{-1}$.
 \label{figure2}
 }
\end{figure}
\end{widetext}

Again, this mechanism, if one were to use it as an explanation of
the magnetoreception phenomena, must be able to explain the
disruptive-field effect.  Figure \ref{figure2}(d) shows the
singlet products for an initial pure triplet state, both with and
without the time-dependent external field switched on
$B_{\mathrm{rf}} = 150$ nT. In this mechanism the sensitivity
itself is still a function of decreasing $\kappa$, as with a zero
$\kappa$ the one-way triplet to singlet transition will completely
dominate the long-time occupations, ultimately giving unit
occupation of the singlet yield.  The effect of the oscillating
field is weaker than that seen for the nuclear hyperfine and
$\Delta g$ mechanisms, but in commonality with those other
mechanisms it occurs for $\kappa < 1\times 10^4$ s$^{-1}$.

We also point out that the ``zero-field effect'' may play a role
in this triplet mechanism. This is a spin-spin interaction effect,
in the presence of spin-orbit coupling, which splits the triplet
states even under zero-external magnetic field. For a radical-pair
of dioxygen and the FAD cofactor in cryptochrome, Hogben {\em et
al.}~\cite{Hogben1} found that the zero-field effect prevented the
normal nuclear hyperfine interaction from being able to explain
the presence of external oscillating fields.  Depending on the
particular radical-pair one is studying this may also apply to our
discussion here, and negatively affect the efficacy of the triplet
mechanism.

\subsubsection{Role of coherence and noise}

The role of quantum coherence in the radical-pair mechanism is
quite subtle. As discussed in the introduction \cite{vedral11},
one can introduce strong pure dephasing which does not affect the
efficacy of the magnetic compass (though other choices of
decoherence do reduce the sensitivity drastically). However, even
pure dephasing alone does effect the ability of the external
oscillating field to disrupt the compass. Some studies
\cite{Cai12} have shown that the environment can improve the
sensitivity of the compass response, with a maximum for an
intermediate level of arbitrary decoherence (both dephasing and
dissipation). For the normal hyperfine mechanism this has been
discussed in great detail in other works
\cite{Cai12,vedral11,Cai2013}. For the $\Delta g$ mechanism one
expects similar results, and some examples were discussed in the
supplementary information of Ref.~\cite{vedral11}.

For the triplet mechanism we make a similar analysis, and
introduce a set of six Lindblad operators, \beq L_i[\rho] =
\Gamma_i \left[\sigma_i^{\alpha} \rho \sigma_i^{\alpha} - \rho
\right] \eeq where $i=1,2;\alpha$ indicates spin $1$ or $2$, and
$\alpha=x,y,z$.  We find, as shown in figure \ref{figure2}(e),
that large `general' decoherence (i.e., equal rates on all six
Lindblad terms), has a detrimental effect on the sensitivity of
the compass. Dephasing in the $\sigma_z$ basis alone reduces but
does not completely eliminate the sensitivity even for relatively
large rates (though this is different from the dephasing in the
eigenbasis which was shown to have no effect on the compass
\cite{vedral11}).  The `increase' in sensitivity with
environmental noise discussed in other works typically arises for
for arbitrary, possibly non-orthogonal, angles in \Eq{sens}, and
we do not discuss that case here.

\subsection{Combined SOC and Nuclear Hyperfine}

\begin{figure}[tbp]
    \includegraphics[width=0.9\columnwidth]{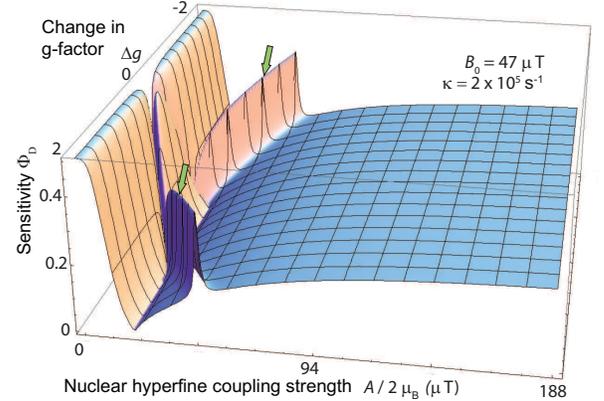}
    \caption{(Color online)
Sensitivity $\Phi_{\rm{D}}$ as a function of $\Delta g$ and
nuclear field strength $A / g \mu_{\rm{B}}$, $B_0=47$ $\mu$T, and
$\kappa=2\times 10^5$ s$^{-1}$. Generally speaking, the combined
effect of spin-orbit coupling and nuclear hyperfine interaction
only reduces the sensitivity, but for small hyperfine coupling
there are two peaks, indicated by the green arrows, where the
combination of both effects can increase the sensitivity.
 \label{figure3}
 }
\end{figure}

One may imagine that in many  realistic situations nuclear
hyperfine and SOC co-exist \cite{Khud, Hogben1}.  In earlier
radical-pair experiments it was generally noted that the
combination of both effects tends to reduce the magnetic
sensitivity of the reactions. Using the analytical results of Cai
{\em et al.}~\cite{Cai12} we can plot the sensitivity as a
function of both $\Delta g$ and nuclear hyperfine coupling
strengths (however only for the hyperfine tensor $(0,0,A_z)$).  We
show the resultant singlet yield sensitivities in figure
\ref{figure3}.  In the parameter regimes where the sensitivity is
highest for each individual mechanism alone we find that the
introduction of the other mechanism generally reduces the
sensitivity.  While there is a large maximum for intermediate
levels of nuclear hyperfine coupling where non-zero SOC increases
the sensitivity (as marked by the green arrows on the figure) if
$\kappa$ is decreased these maxima become increasingly ``narrow'',
and may not represent a robust parameter regime.  These maxima are
broadened somewhat if one performs a maximization of the
sensitivity over all angles in $\Phi(\theta)$, but the behavior
remains qualitatively the same.

\subsection{Distinguishing SOC and Nuclear Hyperfine Mechanisms}

One may be able to distinguish a mechanism based on strong SOC
$\Delta g$ effects from the others (nuclear hyperfine, triplet
mechanism, weak SOC $\Delta g$) by further behavioral experiments
with oscillating fields.  Till now a disruptive effect was found
\cite{Ritz09,Ritz04} when the oscillating field frequency matched
the Zeeman splitting of a radical with $g\approx 2$. Further
experiments may find a disruptive effect when the frequency
matches the energy splitting of the other radical with strong
$\Delta g$. Such disruptive effects are expected to be difficult
to observe in radicals with nuclear hyperfine interactions due to
the large spread of very weak resonances one expects in that case,
when the true nuclear hyperfine interaction is with many nuclear
spins (see the supplementary information of Ref.~\cite{Ritz09} for
a detailed discussion).  Similarly, we expect the weak $\Delta g$
mechanism should not exhibit any additional resonances, and thus
disruption should only occur for the approximately free Zeeman
splitting frequency in that case. A full investigation of the
effect of a disruptive oscillating fields as a function of
frequency, for a range of strong anisotropic $\Delta g$ tensors,
and an analysis of the strength of these effects is a possible
interesting avenue of future work.

Secondly, the weak $\Delta g$ and triplet mechanisms may be
distinguishable from the strong $\Delta g$ and nuclear hyperfine
mechanisms because of the way the disruptive effect of the
oscillating field affects the singlet yield in those cases.  We
have shown in this work, for the weak $\Delta g$ and triplet
mechanisms, that the yield is altered under applied oscillating
fields, but not entirely suppressed. As pointed out earlier, it
has been found that the application of static fields caused a
temporary disruption of the magnetic sense
\cite{wiltstatic,Ritz2011}.  In the standard radical-pair model
static fields were seen to cause a shift of the singlet yield
\cite{ban}, not a suppression.  Logic dictates that if birds were
able to adapt to the disruptive effect of oscillating fields, this
would be an argument in favor of a shift, not a suppression, of
the yield, as seen with static fields, and thus may indicate in
favor of the weak $\Delta g$ or triplet mechanisms. Whether such
adaptation was looked for in experiments so far
\cite{Ritz04,Ritz09} is not clear.

\section{Discussion} \label{CON}

We have shown that spin-orbit effects can, in principle, induce
strong magnetic field sensitivity and satisfy   the experimental
criteria needed as a candidate for magnetoreception. Our main
result is that the $\Delta g$ mechanism becomes more prominent as
the radical recombination rate $\kappa$ is reduced, providing an
explanation for why such a small $\kappa$ might be needed by the
RPM.

If the lifetime of the spin-coherence of the radicals is truly
$100$ $\mu$s then it seems possible that the $\Delta g$ mechanism
could arise and play a role even in organic molecules.  As
discussed in the introduction, the prime candidate so far is the
FAD radical in cryptochrome. However, such a radical is not a good
host of a SOC mediated mechanism as FAD has substantial nuclear
hyperfine interactions \cite{Lau}, which will inevitably dominate
over any weak SOC effects \cite{raey,Kowal04}, even if additional
inorganic radicals are involved \cite{Hogben1}.  This does not
mean that SOC in an alternative radical is not a feasible
mechanism for hosting the magnetic sense of birds, as it is also
difficult to explain the frequency-selective disruptive effect of
oscillating fields \cite{Ritz04,Ritz09} if both radicals contain
strong nuclear hyperfine interactions, and additional difficulties
arise even if FAD forms a radical pair with dioxygen or
superoxide(which have no nuclear hyperfine interactions)
\cite{Hogben1}.

This ultimately leaves open the question of the precise host
radical for the magnetoreception mechanism. Our results would
primarily be relevant for a radical-pair with no, or minimal,
nuclear hyperfine interaction, particularly inorganic radicals
where SOC effects are large, though our results suggest even a
very weak $\Delta g$ anisotropy can induce a magnetic field
sensitivity on the lifetimes apparently demanded by some
behavioral experiments \cite{Ritz04,Ritz09}.  In addition, apart
from providing an alternative route to a magnetically-sensitive
radical pair reaction to describe avian magneto-reception, our
results may help in identifying a broader range of radical pairs
which can be used in artificial weak-magnetic-field sensors
\cite{CaiSun, CaiJ}.

\acknowledgments We thank E. Gauger, C. Ciuti and A.
Smirnov for advice and discussions. We also thank one anonymous
referee for his/her very careful reading of this manuscript and
numerous helpful suggestions. N.~L. acknowledges the hospitality
of the Controlled Quantum Dynamics Group at Imperial College. F.N.
acknowledges partial support from the ARO, RIKEN iTHES Project,
MURI Center for Dynamic Magneto-Optics, JSPS-RFBR Contract No.
12-02-92100, MEXT Kakenhi on Quantum Cybernetics, the
JSPS-FIRST Program, and Grant-in-Aid for Scientific Research (S). S.D.L. acknowledges partial support of the European
Commission under the Marie Curie IEF Program, project BiGExPo

\bibliography{bibliography}

\end{document}